\documentclass[
prl,
twocolumn,
lengthcheck,
superscriptaddress,
]{revtex4-2}

\usepackage{amssymb, amsmath, amsthm, amstext, amsfonts, amsopn, amscd}
\usepackage{mathrsfs, dsfont, mathdots}
\usepackage{bm}

\usepackage{graphicx}
\usepackage[usenames,dvipsnames]{xcolor}

\usepackage[colorlinks=true, linkcolor=black, breaklinks=true, citecolor=blue, urlcolor=red]{hyperref} 
\usepackage{float}
\usepackage{subcaption}
\usepackage{caption}
\usepackage{tikz}
\usepackage{dcolumn}
\usepackage[
mathlines,
pagewise
]{lineno}

\usepackage{todonotes}

\usepackage[
commentmarkup=margin, 
highlightmarkup=footnote,
]{changes}

\makeatletter
\@namedef{Changes@AuthorColor}{red}
\colorlet{Changes@Color}{red}
\makeatother

\usetikzlibrary{decorations.markings}

\theoremstyle{definition}

\theoremstyle{remark}

\DeclareMathOperator{\1}{\mathds{1}}
\DeclareMathOperator{\tr}{tr}
\DeclareMathOperator{\Tr}{Tr}

\newcommand{\fA}{\mathfrak A}

\newcommand{\bC}{\mathbb C}

\newcommand{\cE}{\mathcal E}

\newcommand{\cH}{\mathcal H}

\newcommand{\R}{\mathds{R}}
\newcommand{\cS}{\mathcal{S}}
\newcommand{\T}{\mathds T}

\newcommand{\varep}{\varepsilon}

\newcommand{\Z}{\mathds{Z}}

\begin{document}


\tikzset{->-/.style={decoration={
  markings,
  mark=at position #1 with {\arrow{latex}}},postaction={decorate}}}

\title{Operator-algebraic renormalization and wavelets}
\author{Alexander Stottmeister}
\affiliation{Institute of Theoretical Physics, University of Hannover, Appelstra\ss e 2, 30167 Hannover, Germany}
\author{Vincenzo Morinelli}
\affiliation{Department of Mathematics, FAU Erlangen-N\"urnberg, Cauerstra\ss e 11,  91058 Erlangen, Germany}
\author{Gerardo Morsella}
\author{Yoh Tanimoto}
\affiliation{Department of Mathematics, University of Rome ``Tor Vergata'', Via della Ricerca Scientifica 1, 00133 Roma, Italy}

\date{\today}

\begin{abstract}
We report on a rigorous operator-algebraic renormalization group scheme and construct the free field with a continuous action of translations as the scaling limit of Hamiltonian lattice systems using wavelet theory.
A renormalization group step is determined by the scaling equation identifying lattice observables with the continuum field smeared by
compactly supported wavelets.
Causality follows from Lieb-Robinson bounds for harmonic lattice systems.
The scheme is related with the multi-scale entanglement renormalization ansatz and augments the semi-continuum limit of quantum systems.
\end{abstract}

\maketitle

\section{Introduction}
Lattice regularization is a standard procedure to define continuum quantum field theories \cite{FernandezRandomWalksCritical} which has led to extraordinary results in the ab-initio determination of the Hadron mass spectrum \cite{DuerrAbInitioDetermination} and may serve as a starting point for the quantum simulation of quantum field theories \cite{JordanQuantumAlgorithmsFor}. While interacting models have been rigorously constructed in the classical works of Glimm-Jaffe and others \cite{GlimmQuantumFieldTheory} , the lattice and continuum theories
are often related indirectly in terms of correlation functions.

A recent attempt to build a continuum conformal field theory (CFT) by embedding a quantum spin chain from coarser to finer lattices, coined the semi-continuum limit and inspired by block-spin renormalization, resulted in a discontinuous action
of symmetries, even the translations \cite{JonesANoGo, JonesScaleInvariantTransfer, KlieschContinuumLimitsOf, OsborneQuantumFieldsFor}.
Here, we explain how this deficiency can be remedied by utilizing an observable-based, i.e.~operator-algebraic, approach to the Wilson-Kadanoff renormalization group (RG) \cite{KadanoffScalingLawsFor, WilsonTheRenormalizationGroupKondo, FisherRenormalizationGroupTheory} for lattice field theories \cite{BrothierConstructionsOfConformal, BrothierAnOperatorAlgebraic}. As an important, instructive example \cite{WhiteRealSpaceQuantum1,WhiteDensityMatrixFormulation}, we construct the massive continuum free field with its continuous action of spacetime translations via the scaling limit of lattice systems in their ground states approaching the unstable, massless fix point (\cite{MorinelliScalingLimitsOf} for details and proofs). More recently, the presented method has been extended to CFTs based on free fermions \cite{OsborneCFTapprox} invoking the Koo-Saleur formula \cite{KooRepresentationsOfThe}.

Our RG is defined in terms of compactly supported, regular wavelets \cite{DaubechiesTenLecturesOn} allowing for simultaneous control of locality properties in real and momentum space.
\begin{figure}[ht]
\scalebox{0.9}{
\includegraphics{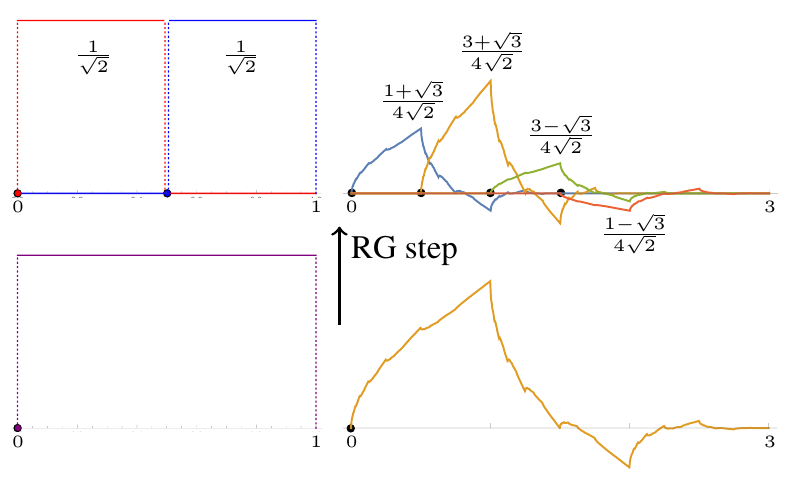}
}
\caption{\small Illustration of the decomposition of lattice sites for $d=1$ by an RG step determined by the scaling equation \eqref{eq:scalingeq}: On the left: The block-spin RG and its weights. On the right: The wavelet-based RG with weights determined by the low-pass filter of Daubechies' D4 scaling function.}
\label{fig:D4}
\end{figure}
We take inspiration from renormalization in classical systems \cite{BattleWaveletsAndRenormalization} and use a scaling function and its multiresolution analysis to define a RG step: While block-spin renormalization would correspond to a step function, we use a Daubechies scaling function (see Figure \ref{fig:D4}), cf.~\cite{BrennenMultiscaleQuantumSimulation, EvenblyEntanglementRenormalizationIn}. Thereby we avoid the obstacles encountered in \cite{JonesANoGo, KlieschContinuumLimitsOf, OsborneQuantumFieldsFor} to implement continuous symmetries in the scaling limit, cf.~\cite{ZiniConformalFieldTheories}. Mapping observables from coarser to finer lattices results in a real-space RG dual to coarse graining the Hamiltonian or density matrices, e.g.~the density matrix renormalization group (DMRG) \cite{WhiteDensityMatrixFormulation, WhiteDensityMatrixAlgorithms, SchollwoeckTheDensityMatrix}. Our method applies in all dimensions as we explicitly demonstrate for scalar lattice fields. Moreover, our approach yields a rigorous proof that spacetime locality
(in the sense of the Haag-Kastler axioms \cite{HaagLocalQuantumPhysics})
in the continuum follows from Lieb-Robinson bounds \cite{LiebTheFiniteGroup, CramerLocalityOfDynamics, OsborneContinuumLimitsOf, NachtergaeleLRBoundsHarmonic, NachtergaeleQuasiLocalityBounds}. 

As real-space RG schemes have received rapidly growing interest in recent years, especially in the context of tensor networks \cite{CiracRenormalizationAndTensor} and the multi-scale entanglement renormalization ansatz (MERA) \cite{VidalAClassOf, EvenblyAlgorithmsForEntanglement, PfeiferEntanglementRenormalizationScale}, we show as an important application that our approach yields a rigorous analytic MERA in any dimension $d$ which is not restricted to critical (massless) models \cite{EvenblyEntanglementRenormalizationAnd, EvenblyRepresentationAndDesign}.
The discrete dimension of the $d+1$-dimensional tensor network of the MERA is identified with the sequence of scales the given quantum system is observed at.

The letter is organized as follows. First, we outline our general renormalization scheme. Then, we apply it to lattice scalar fields by constructing explicit renormalization maps in terms of compactly supported wavelets, and we discuss the connection with the MERA. Finally, in the example of the free scalar field, we show that imposing a suitable renormalization condition on lattice ground states at different scales, we fully recover the continuum massive field in the scaling limit including the action of spacetime translations. The letter closes with an outlook on possible future developments.
\section{Operator-algebraic renormalization}
\begin{figure}[ht]
\scalebox{1}{
\includegraphics{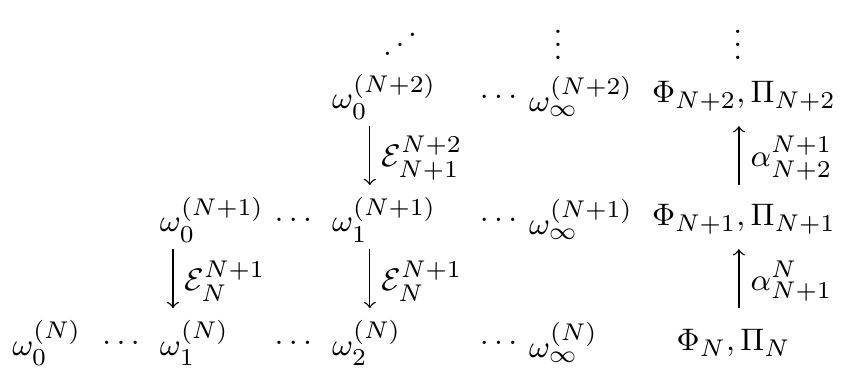}
%
%
%
%
%
}
\caption{\small Wilson's triangle of renormalization: Vertical lines represent renormalization steps, either by coarse graining states ($\cE$'s) or by refining fields ($\alpha$'s). Horizontal lines represent sequences of renormalized states considered on the algebra generated by fields and momenta at a fixed scale (right column).}
\label{fig:statetrianglerg}
\end{figure}

As discussed in \cite{BrothierAnOperatorAlgebraic}, the RG approach to the lattice approximation of continuum theories can be rephrased in terms of observables, that is operator algebras, as follows. We fix a family of lattices $\Lambda_{N}$ in $\R^{d}$ with lattice constant $\varep_{N} = 2^{-N}\varep$,
and consider a sequence of Hamiltonian quantum systems $\{\fA_{N},\cH_{N},H^{(N)}_{0}\}$ indexed by the scale $N$. At each scale $N$, we have an algebra of observables $\fA_{N}$ generated by (bounded functions of) basic time-zero lattice fields $\Phi_N(x)$, their momenta $\Pi_{N}(x)$, and a Hamiltonian $H^{(N)}_{0}$ both acting on the Hilbert space $\cH_{N}$.
The quantum state at each scale is initially given by a density matrix $\rho^{(N)}_{0}$, e.g.~in terms of a Hamiltonian: $\rho^{(N)}_{0}\!=\!(Z^{(N)}_{0})^{-1}e^{-H^{(N)}_{0}}$.
The RG connects systems at different scales via (coarse graining) quantum operations, mapping density matrices on the finer system to the coarser system
\begin{align}
\label{eq:densityrg}
\cE^{N+M}_{N}\!(\rho^{(N+M)}_{0}) &\!=\!\rho^{(N)}_{M}, & \cE^{N+1}_{N}\circ\cE^{N+2}_{N+1} & \!=\!\cE^{N+2}_{N},
\end{align}
where $\rho^{(N)}_{M}$ corresponds to the ($M$ times) renormalized Hamiltonian $H^{(N)}_M$ at scale $N$. Because quantum states $\rho$ are positive, linear maps $\omega : \fA_N \to \bC$, by $\omega(A) = \tr(\rho A)$, and the field correlation functions are given by $\langle \Phi_N(x)\dots \Pi_N(y)\rangle^{(N)} := \omega^{(N)}(\Phi_N(x)\dots \Pi(y))$, we can state \eqref{eq:densityrg} as:
\begin{align}
\label{eq:staterg}
\cE^{N+M}_{N}(\omega^{(N+M)}_{0}) & = \omega^{(N+M)}_{0}\circ\alpha^{N}_{N+M} = \omega^{(N)}_{M},
\end{align}
where $\alpha^{N}_{N+M}:\fA_{N}\rightarrow\fA_{N+M}$ is the dual of $\cE^{N+M}_{N}$ (the ascending superoperators \cite{EvenblyAlgorithmsForEntanglement}). $\omega^{(N)}_{0}$ and $\omega^{(N)}_{M}$ characterize the initial and renormalized states on $\fA_N$ corresponding to $\rho^{(N)}_{0}$ and $\rho^{(N)}_{M}$. We call the collection $\alpha^{N}_{N+M}$, the \textit{scaling maps} or \textit{renormalization group}. The structure is neatly summarized by an adaptation of Wilson's \textit{triangle of renormalization} \cite[p. 790]{WilsonTheRenormalizationGroupKondo} in Figure \ref{fig:statetrianglerg}.
If the limit $\omega_\infty^{(N)} := \lim_{M \to \infty} \omega^{(N)}_M$ exists (in a suitable sense), the sequence $\omega_\infty^{(N)}$, called the \emph{scaling limit} of the inital states  $\omega^{(N)}_{0}$, is stable under coarse graining:
\begin{align}
\label{eq:staterglimit}
\cE^{N+M}_{N}(\omega^{(N+M)}_{\infty}) & = \omega^{(N)}_{\infty}, & N&\!<\!N'.
\end{align}
Employing operator-algebraic techniques (see \cite{MorinelliScalingLimitsOf} for details), we obtain a Hilbert space $\cH_\infty$ and an algebra $\fA_\infty$ generated by continuum fields $\Phi, \Pi$, acting on it. Following \cite{JonesANoGo, JonesScaleInvariantTransfer, MilstedQuantumYangMills, BrothierConstructionsOfConformal, BrothierAnOperatorAlgebraic} we call $\fA_\infty$ the \textit{semi-continuum limit}, see also \cite{KijowskiSymplecticGeometryAnd, KijowskiAModificationOf}. Moreover, we have isometries $V^{N}_\infty : \cH_N \to \cH_\infty$ and a state $\Omega\in\cH_{\infty}$ realizing the correlations of the scaling limit $\omega\!=\!\langle\Omega,\!\ .\!\ \Omega\rangle$. The finite-scale fields $\Phi_{N},\Pi_{N}$
are embedded in the continuum fields $\Phi,\Pi$ through $\alpha^{N}_{\infty}:\fA_{N}\rightarrow\fA_{\infty}$:
\begin{align}
\label{eq:statelimiso}
\alpha^{N}_{\infty}(\Phi_{N}(x))V^{N}_{\infty} & = V^{N}_{\infty}\Phi_{N}(x), & \omega^{(N)}_{\infty} &\!=\!\omega\circ\alpha^{N}_{\infty} .
\end{align}
\section{Wavelets and the scalar field}
\begin{figure}[ht]
\scalebox{0.9}{
\includegraphics{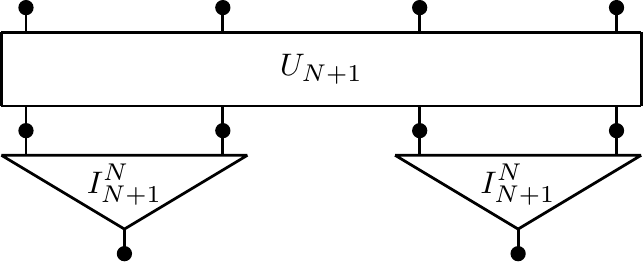}
%
%
%
%
%
}
\caption{\small Illustration of the analytic MERA in $d=1$ induced by the wavelet scaling maps. From bottom to top: the first layer represents the isometric embedding $I_{N+1}^N$ and the second layer represents the action of the (dis)entangler $U_{N+1}$ at scale $N+1$.}
\label{fig:MERA}
\end{figure}

We now apply the above framework to lattice scalar fields, setting up a specific renormalization scheme involving compactly supported wavelets \cite{DaubechiesTenLecturesOn, MeyerWaveletsAndOperators}.
To avoid infrared divergence at finite scale, we take lattices $\Lambda_{N}\!=\!\varep_{N}\{-L_{N},...,L_{N}-1\}^{d}$ representing a discretization of the torus $[-L,L)^{d}\!=\!\T_{L}^{d}$ (periodic boundary conditions, $L_{N}\!\equiv\!-L_{N}$, with $\varep_{N} L_{N}\!=\!L$ fixed). We denote by $\Gamma_{N}\!=\!\tfrac{\pi}{L}\{-L_{N},...,L_{N}-1\}^{d}$ the dual momentum space lattices. The kinematical setup of the lattice scalar field systems is given by the Fock space $\cH_N$, built from the action of momentum-space creation and annihilation operators $a_N(k), a^{\dag}_N(k)$ on the vacuum vector $\Omega_N$ subject to the canonical commutation relations (CCR), $
[a_N(k), a^{\dag}_N(l)]\!=\!(2L_N)^{d} \delta_{k,l}$, and by the algebra $\fA_N$ generated by the local (dimensionless) canonical lattice field for $x \in\Lambda_N$:
\begin{align*}
\Phi_N(x) &= \tfrac{1}{\sqrt{2}(2L_{N})^{d}}\sum_{k \in \Gamma_N} [a^{\dag}_N(k) e^{-ikx}+a_N(k) e^{ikx}],
\end{align*}
and its momentum (with a similar formula) satisfying: $[\Phi_N(x),\Pi_N(y)]\!=\!i\delta_{x,y}$. The scaling maps $\alpha^N_{N'} : \fA_N \to \fA_{N'}$ are the most important input in our framework determining the existence and structure of the continuum limit. Our choice using wavelets is motivated by the block-spin case and its locality properties in real space corresponding to the smearing of continuum fields with the simplest member of the Daubechies' wavelet, the \textit{Haar wavelet} $\chi_{[0,1)}$ (see Figure \ref{fig:D4}). But, as the approximation of momenta requires higher regularity, the latter does not suffice as explained below.
\paragraph{Scaling maps from a scaling function.}
We consider an orthonormal scaling function $s$ that satisfies the \textit{scaling equation} \cite{MallatMultiresolutionApproximationsAnd, MeyerPrincipeDIncertitude, DaubechiesTenLecturesOn}:
\begin{align}\label{eq:scalingeq}
s(x) &\!=\!\sum_{n\in\Z^{d}}h_{n}2^{\frac{d}{2}}s(2x-n),
\end{align}
such that its integer translates $s(\cdot-\!n)$ are orthonormal. To build local operators, we further take $s$ compactly supported and normalized by $\hat{s}(0)=1$. Such an $s$ generates an orthonormal, compactly supported wavelet basis in $L^{2}(\R^d)$, and the sum \eqref{eq:scalingeq} is necessarily finite ($h_{n}$ is a finite low-pass filter \cite{DaubechiesTenLecturesOn}). We denote by $s^{(\varep)}_{x}\!=\!\varep^{-\frac{d}{2}}s(\varep^{-1}(\cdot-x)\!)$  the scaling function localized near $x\!\in\!\varep\Z^{d}$ at length scale $\varep$, periodized on the torus $\T_{L}^{d}$. With the scaling relation \eqref{eq:scalingeq} in mind,
we define $\alpha^{N}_{N+1}$ using the low-pass filter $h_n$:
\begin{align}
\label{eq:oneparticlescaling}
\alpha^{N}_{N+1}(\Phi_N(x)) &\!=\!2^{-\frac{1}{2}}\sum_{n\in\Z^{d}}h_{n}\Phi_{N+1}(x+n\varep_{N+1}),
\end{align}
and similarly for $\Pi_N$. Now, the associated semi-continuum limit algebra $\fA_{\infty}$ can be identified with the algebra generated by continuum fields smeared with the functions $s^{(\varep_N)}_x$ over all scales $N$: The map,
\begin{align}
\label{eq:contembedd}
\Phi_{N}(x) & \mapsto \alpha^{N}_{\infty}(\Phi_{N}(x))\!=\!\varep_{N}^{-\frac{1}{2}}\!\int\!\!dy \,\Phi(y)s^{(\varep_N)}_x(y),
\end{align}
identifies the lattice fields at scale $N$ with the continuum fields smeared with $s^{(\varep_N)}_x$ (and analogously for $\Pi_{N}(x)$). The RG elements $\alpha^{N}_{N'}$ defined by \eqref{eq:oneparticlescaling} have two intriguing properties: First, the lattice field $\Phi^{(N)}(x)$ at one scale is decomposed into a linear combination of the fields at the successive scale. Second, the embedding \eqref{eq:contembedd} into the continuum field theory is compatible with this decomposition, $\alpha^{N+1}_{\infty}\circ\alpha^{N}_{N+1}\!=\!\alpha^{N}_{\infty}$, realizing the correct CCR:
\begin{align*}
[\alpha^{N}_{\infty}\!(\Phi_{N}(x)\!),\alpha^{N}_{\infty}\!(\Pi_{N}(y)\!)] & \!=\![\Phi(s^{(\varep_N)}_{x}), \Pi(s^{(\varep_N)}_{y})]\!=\!i\delta_{x,y}.
\end{align*}
Furthermore, we have $\Phi(s^{(\varep_N)}_{x})\!=\!\sum_{n\in \varep_N}\!h_n\Phi(s^{(\varep_{N+1})}_{x-n\varep_{N+1}})$ (linearity and \eqref{eq:scalingeq}) with an analogous formula for $\Pi$. This means that the lattice fields and their realization in terms of the continuum field have the same algebraic structure. 
\paragraph{Concrete choice of a scaling function.}
The simplest scaling function, $\chi_{[0,1)}$, corresponds to the block-spin renormalization \eqref{eq:oneparticlescaling} (see Figure \ref{fig:D4}). By taking a more regular scaling function, e.g.~$_Ks$ with $K\ge 2$ of Daubechies' D2K wavelet family, we achieve that the smeared continuum momentum $\Pi(_Ks^{(\varep_{N})}_{x})$ is a well-defined operator (technically $s$ needs to be in the Sobolev space $H^\frac{1}{2}$). In addition, the compact support of $_Ks$ leads to locality in real space, i.e.~the lattice fields $\Phi_{N}(x), \Pi_{N}(x)$ can be used to approximate local operators in the continuum because $\Phi(s^{(\varep_N)}_{x}), \Pi(s^{(\varep_N)}_{x})$ are spatially localized in compact regions. In comparison with the block-spin renormalization we trade some locality (the support of the Daubechies scaling function $_Ks$ is larger than the support of $\chi_{[0,1)}$) for higher regularity improving approximations. With this price, we gain the continuum realization of $\Pi_{N}(x)$, and we recover the correlation functions and space-time symmetries (translations) in the scaling limit (see below).
\paragraph{Connection with multi-scale entanglement renormalization.} 
Considering the embedding $I^{N}_{N+1}(\Phi_{N}(x)) = 2^{-\frac{1}{2}}\Phi_{N+1}(x)$ resulting from identifying $\Lambda_{N}$ as a sublattice of $\Lambda_{N+1}$, and the Bogoliubov unitary,
\begin{align}
\label{eq:symplecticrot}
U_{N+1}\Phi_{N+1}(x) &\!=\!\sum_{n\in\Z^{d}}\!h_{n}\Phi_{N+1}(x\!+\!n\varep_{N+1})U_{N+1},
\end{align}
implementing the redistribution of field values according to the low-pass filter $h_{n}$, the scaling map $\alpha^{N}_{N+1}$ decomposes into MERA form \cite{VidalAClassOf, EvenblyAlgorithmsForEntanglement, PfeiferEntanglementRenormalizationScale, MilstedQuantumYangMills, BrothierAnOperatorAlgebraic}: 
\begin{align}
\label{eq:meraalg}
\!\!\!\alpha^{N}_{N+1}(\!\ \cdot\!\ ) & \!=\!U_{N+1}(\!\ \cdot\!\ \!\otimes\!\1_{N+1\setminus N}\!)U_{N+1}^{*},
\end{align}
Here, $\!\ \cdot\!\ \otimes\1_{N+1\setminus N}$ is the tensor product with the identity on the ancillary Fock space, $\cH_{N+1}\!=\!\cH_{N}\otimes\cH^{(a)}_{N+1}$, and the dual quantum channel $\cE^{N+1}_{N}\!=\!\Tr_{\cH^{(a)}_{N+1}}(U_{N+1}^{*}(\!\ \cdot\!\ )U_{N+1})$ is given by a twisted partial trace on the ancillary. From \eqref{eq:meraalg}, we find that $U_{N+1}$ serves as MERA disentangler recovered from the isometries, $V^{N}_{N+1}:\cH_{N}\rightarrow\cH_{N+1}$, between Fock spaces resulting from coarse-graining stability \eqref{eq:staterglimit}:
\begin{align}
\label{eq:meragns}
\Omega^{(N+1)}_{\infty} &\!=\!V^{N}_{N+1}\Omega^{(N)}_{\infty},
\end{align}
where $\Omega^{(N)}_{\infty}$ is the vector implementing the scaling limit $\omega^{(N)}_{\infty}$ at scale $N$. The embedding into the continuum Hilbert space $\cH_{\infty}$ can be explicitly computed from \eqref{eq:statelimiso}. Summarizing, we observe that one layer of MERA isometries and disentanglers is recovered from $\alpha^{N}_{N+1}$ and the scaling limit $\omega^{(N)}_{\infty}$. This structure is further elucidated by the action of the isometries $V^{N}_{N+1}$ on coherent or Glauber states, $c_{N}(f,g)\!= \!e^{i(\Phi_{N}(f)+\Pi_{N}(g))}\Omega^{(N)}_{\infty}$, using the identification \eqref{eq:contembedd} (see Figure \ref{fig:MERA}). 
In this sense, our operator-algebraic RG scheme produces an analytic MERA. Specifically, the scaling limits of free lattice ground states, which we construct below, exhibit a structure similar to an analytic MERA in arbitrary dimensions and off criticality \cite{EvenblyEntanglementRenormalizationAnd, HaegemanRigorousFreeFermion, WitteveenQuantumCircuitApproximations, WitteveenWaveletConstructionOf}.
\section{Scaling limits of harmonic lattice systems}
We are now in a position to apply the RG $\alpha^{N}_{N'}$ defined by \eqref{eq:oneparticlescaling} to find the ground-state scaling limits of the free lattice Hamiltonian on $\cH_{N}$:
\begin{align}
\label{eq:freeham}
H^{(N)}_{0} &\!\!=\!\varep_{N}^{-1}\hspace{-0.1cm}\Bigg(\hspace{-0.1cm}\tfrac{1}{2}\!\hspace{-0.15cm}\sum_{x\in\Lambda_{N}}\hspace{-0.17cm}\!\big(\Pi_{N|x}^{2}\hspace{-0.15cm}+\!\mu^{2}_{N}\Phi_{N|x}^{2}\!\big)\hspace{-0.1cm}-\!\!\hspace{-0.45cm}\sum_{\langle x,y\rangle\subset\Lambda_{N}}\hspace{-0.42cm}\Phi_{N|x}\Phi_{N|y}\hspace{-0.15cm}\Bigg),
\end{align}
where $\mu_{N}\geq2d$ is a ``mass'' parameter. The ground state $\Omega^{(N)}_{0}$ of $H^{(N)}_{0}$ can be encoded into the expectation $\omega^{(N)}_{0}$ on $\fA_N$ determined by the two-point functions:
\begin{align}
\label{eq:freegroundPhi}
\omega^{(N)}_{0}\!(\Phi_N(x) \Phi_N(y)\!) &\!=\!\! \tfrac1{(2L_N)^d}\!\!\!\sum_{k \in \Gamma_N}\!\!\! \tfrac{1}{2\varep_{N}\gamma_{\mu_N}\!(k)}e^{ik(x-y)},
\end{align}
with the dispersion relation $\gamma_{\mu_{N}}^{2}(k)\!=\!\varep_{N}^{-2}(\mu_{N}^{2}\!-\!2d)\!+\!2\epsilon_{N}^{-2}\sum^{d}_{j=1}(1\!-\!\cos(\varep_{N}k_{j})\!)$, and analogous formulae for $\omega^{(N)}_{0}\!(\Phi_N(x) \Pi_N(y)\!)$ and $\omega^{(N)}_{0}\!(\Pi_N(x) \Pi_N(y)\!)$, the latter being most singular.
\paragraph{Scaling limit of the ground states.}
We choose \eqref{eq:freegroundPhi} as our initial states to generate a sequence of renormalized states $\omega^{(N)}_{M}$ at each scale $N$
(Figure \ref{fig:statetrianglerg}). To avoid the RG-fixed points $\mu_{N}^{2}\!=\!2d$ (massless, unstable) and $\mu_{N}^{2}\!=\!\infty$ (ultralocal, stable) and hit the unstable manifold of the relevant $\Phi^{2}$-operator, we impose the \textit{renormalization condition},
\begin{align}
\label{eq:rencon}
\lim_{N\rightarrow\infty}\varep^{-2}_{N}(\mu_{N}^{2}-2d) & = m^{2},
\end{align}
for some $m\!>\!0$. This leads to the massive continuum dispersion, $\lim_{M\rightarrow\infty}\gamma_{\mu_{N+M}}(k)^{2}\!=\!m^{2}+k^{2}\!=\!\gamma_{m}(k)^{2}$, and the scaling limit (using \eqref{eq:oneparticlescaling} \& \eqref{eq:freegroundPhi}, and similar for $\Pi_{N}$):
\begin{align}
\label{eq:freegroundlimitPhi}
\omega^{(N)}_{m,\infty}\!(\Phi_N(x) \Phi_N(y)\!) &\!=\! \tfrac1{(2L)^d}\!\!\!\sum_{k \in \Gamma_\infty}\!\!\! \tfrac{|\hat{s}^{(\varep_N)}(k)|^2}{2\varep_{N}\gamma_{m}(k)} e^{ik(x-y)},
\end{align}
where $\Gamma_\infty\!=\!\tfrac{\pi}{L}\Z^{d}$ is the momentum space of the torus $\T^{d}_{L}$.
Since the two-point function of the momentum $\Pi_N$ is the most singular, the limit states are well defined for scaling functions with sufficient momentum-space decay, which holds for scaling functions $_{K}s$, $K\!\geq\!2$, built from Daubechies' D2K wavelet family \cite{DaubechiesTenLecturesOn}.
Formulas \eqref{eq:freegroundlimitPhi}, multiplied by $\varep_N$, $\varep_N^{-1}$ respectively, agree with the two point functions of the usual continuum mass-$m$ ground state in finite volume $L$ of the continuum smeared field operators $\Phi(s^{(\varep_N)}_x)$, $\Pi(s^{(\varep_N)}_x)$.
Therefore, the semi-continuum limit algebra $\fA_\infty$ can be identified with a subalgebra of the algebra $\fA_{m,L}$ generated by the massive continuum free field ($m>0$) on $\T^d_L$, acting on the usual continuum Fock space. Because of localization and completeness of the wavelet basis associated with the scaling function $s$
\cite{MeyerWaveletsAndOperators, DaubechiesTenLecturesOn}, all field operators $\Phi(f)$, $\Pi(g)$ smeared with smooth compactly supported functions can be approximated, in an appropriate sense, by operators from $\fA_\infty$.
\paragraph{Translations, dynamics, locality and Lieb-Robinson bounds.}
Our construction provides an explicit method to circumvent the no-go results of \cite{JonesANoGo, KlieschContinuumLimitsOf} concerning the implementation of continuous symmetries.
In particular, the continuous extension of spatial translations by discrete vectors $a \in \bigcup_N \Lambda_N$ (dyadic translations as enforced by the dyadic lattice refinements) acting on $\fA_\infty$ to translations by arbitrary vectors $a \in \T^d_L$ is a consequence of the manifest continuous translations invariance of the two-point function \eqref{eq:freegroundlimitPhi}, and the generators of translations are the usual momentum operators. The thermodynamical limit of \eqref{eq:freegroundlimitPhi}, $L\rightarrow\infty$, exists by a Riemann-sum argument and yields the two-point functions of the free, massive vacuum in infinite volume (see \cite{MorinelliScalingLimitsOf}), which is fully Poincar\'e invariant.
Let us also explicitly address the convergence of the lattice dynamics generated by the Hamiltonian $H^{(N)}_0$ of~\eqref{eq:freeham} to their continuum limit:
From $\gamma_{\mu_{N}}\!\rightarrow\!\gamma_{m}$ we deduce
\begin{align*}
V^{N'}_\infty\!e^{itH_0^{(N')}}\!\!\alpha^N_{N'}(\Phi_N(x)\!)\Omega^{(N)}_{\infty} & \!\stackrel{N'\rightarrow\infty}{\rightarrow}\! e^{itH} V^N_\infty\Phi_N(x) \Omega^{(N)}_{\infty},
\end{align*}
and similarly for $\Pi_N$, uniformly on bounded intervals of $t \in \R$, with the free continuum Hamiltonian $H$ on the torus $\T^d_L$.
Since $\gamma_{m}$ is the free, massive relativistic dispersion relation, we know that the dynamics generated by $H$ has propagation speed $c\!=\!1$ and, thus, the scaling limit theory satisfies Einstein causality, i.e., $e^{itH}\alpha^N_\infty(\Phi_N(x))e^{-itH}$ and $e^{isH}\alpha^N_\infty(\Phi_N(x))e^{-isH}$ commute if the support of $s^{(\varep_N)}_x$ at time $t$ and the support of $s^{(\varep_N)}_y$ at time $s$ are spacelike separated on the torus.
A more lattice-intrinsic and model-independent way to conclude recovery of causality in the scaling limit is via Lieb-Robinson bounds \cite{CramerLocalityOfDynamics, OsborneContinuumLimitsOf}.
Considering the extension of the finite-scale time translations $\sigma^{(N)}_{t}\!=\!e^{itH_0^{(N)}}(\cdot)e^{-itH_0^{(N)}}$ to $\fA_\infty$ by \eqref{eq:meraalg}, 
said bounds for harmonic lattice systems \cite{NachtergaeleLRBoundsHarmonic} imply:
\begin{align}
\label{eq:LRlocal}
\lim_{N\rightarrow\infty}\Big[\sigma^{(N)}_{t}(A),B\Big]& \!=\!0,
\end{align}
exponentially fast and uniformly for $|t|\!\leq\!T$ with (bounded) $A, B \in \fA_\infty$ localized in sets $\cS_A, \cS_B \subset \T^d_L$ such that $\textup{dist}(x,\cS_A) \geq c'T$ for all $x \in \cS_B$, for some $c' > 1$.
Because $c'\!>\!1$, the causality implied by \eqref{eq:LRlocal} is not strict likely due to a non-optimal bound on the Lieb-Robinson velocity \cite{CramerLocalityOfDynamics}. 
Another important feature of our approximation of dynamics (or symmetries in general) is the possibility for uniform error bounds in time and within a fixed range of field and momentum amplitudes at a given scale $N$: For the free continuum time evolution $\sigma_{t}\!=\!e^{itH}(\cdot)e^{-itH}$ we have \cite{MorinelliScalingLimitsOf}:
\begin{align}
\label{eq:dynapprox}
\|(\sigma^{(N')}_{t}\!\!\!-\!\sigma_{t})(A)\psi\| & \!\!\leq\!\! C\!\!\sup_{k\in\Gamma_{\infty}}\!\!\!\Big(\!\tfrac{\gamma_{m}(k)^{\!\frac{1}{2}}\!|\gamma_{\mu_{N'}}\!(k)-\gamma_{m}(k)|}{(1+\varep_{N}|k|)^{\delta}}\!\Big),
\end{align}
for exponentials $A\!=\!\alpha^{N}_{\infty}(e^{i(\Phi_{N}(x)+\Pi_{N}(y))})$ of fields and momenta on coherent states $\psi\!=\!c(\varep_{N}^{-\frac{1}{2}}s^{(\varep_{N})}_{u},\varep_{N}^{\frac{1}{2}}s^{(\varep_{N})}_{v})$ at scale $N$. $C$ only depends on $N, \varep_{N}, m, T$ for $|t|\leq T$, and $s$.
While the specific form of these bounds reflects the free-field situation, our general method to obtain such uniform bounds at fixed approximation scale $N$ is not restricted to this situation (cf. conclusion).
\section{Conclusions and outlook}
Our results show that the existence and properties of continuum limits depend decisively on the choice of a renormalization scheme.
Correctly choosing the initial states allows us to reconstruct the continuum field theory from the lattice approximation through the semi-continuum limit.
For the free massive scalar field, our renormalization scheme, given by compactly supported wavelets, yields continuous spacetime translations,
avoiding the apparent no-go results stated in \cite{JonesANoGo, KlieschContinuumLimitsOf}.
Obtaining a similar convergence statement for Lorentz transformations or even conformal transformations requires further work \cite{OsborneCFTapprox}.
Apart from the question of approximation of symmetries, our method proves (\eqref{eq:freegroundlimitPhi} and \eqref{eq:dynapprox}) that time-dependent and spatially translated correlation functions of the continuum field theory for any insertions of fields and momenta, $A_{N}\!=\!\Phi_N(x_{1})...\Pi_N(x_{n})$ and $B_{N}\!=\!\Phi_N(x_{n+1})...\Pi_N(x_{n+m})$, at any scale $N$ are approximated by the correlation functions of the lattice models (suppressing scaling maps $\alpha^{N}_{N'}$, $\alpha^{N}_{\infty}$):
\begin{align}
\label{eq:corapprox}
|\omega^{(N')}_{0}\!(A_{N}\sigma^{(N')}_{(t,x)}(B_{N})\!)\!-\!\omega(A_{N}\sigma_{(t,x)}\!(B_{N})\!)\!| & \!\!\stackrel{N'\rightarrow\infty}{\rightarrow}\!0,
\end{align}
where $\sigma_{(t,x)}$ and $\sigma^{(N')}_{(t,x)}$ are the continuum respectively discrete spacetime translations for $(t,x)\in\R\times\Lambda_{N}$. We point out that the convergence in \eqref{eq:corapprox} only mildly depends on the choice of scaling function $s$ (requiring sufficient regularity). This presents a significant conceptual and presumably computational difference in comparison with a related construction using wavelet theory \cite{WitteveenWaveletConstructionOf} focusing on locality in one-particle space and relying on a continuous adaptation of the choice of scaling function to achieve a given accuracy goal for the approximation of equal-time correlation function similar to \eqref{eq:corapprox}.
An application of the wavelet method to (free) lattice fermions has lead to similar results as those presented here \cite{OsborneCFTapprox, OsborneCFTsim}.
Our general framework can also include interacting lattice systems, e.g.~$\Phi^4$-models,
although we will need approximations by analytical and numerical expansion or perturbative methods
\cite{BorgsConfinementDeconfinementAnd, SchollwoeckTheDensityMatrix, BrydgesLecturesOnThe}.
Moreover, Lieb-Robinson bounds for anharmonic lattice systems \cite{NachtergaeleQuasiLocalityBounds} offer a possibility to obtain spacetime locality directly from the lattice \cite{CramerLocalityOfDynamics, OsborneContinuumLimitsOf}. In view of the classical results by Glimm-Jaffe and others \cite{GlimmQuantumFieldTheory} on $P(\Phi)$-models in $d\!=\!1$, our method is directly applicable to those using a low-pass filter implementing momentum-space cutoffs \cite{MorinelliScalingLimitsOf} thereby providing the same regularized continuum fields as in \cite{GlimmJaffe1}, and we expect a possible extension to the wavelet setting. Therefore, it would be interesting whether the convergence to the scaling limit can be shown exploiting the results in \cite{GlimmJaffe3} supplemented by explicit error bounds similar to \eqref{eq:dynapprox}.
\begin{acknowledgments}
Helpful discussions with T. Osborne, A. Abdesselam and M. Fr{\"o}b are acknowledged by AS. We would also like to thank the unknown referees for their careful consideration of our manuscript thereby improving the clarity of the presentation.
VM and GM are partially supported by the European Research Council Advanced Grant 669240 QUEST. AS was supported by the Humboldt Foundation through a Feodor Lynen Return Fellowship.
VM was titolare di un Assegno di Ricerca dell'Istituto Nazionale di Alta Matematica (INdAM fellowship).
YT is supported by the Programma per giovani ricercatori, anno 2014 ``Rita Levi Montalcini''
of the Italian Ministry of Education, University and Research.
VM, GM and YT also acknowledge the MIUR Excellence Department Project awarded to the Department of Mathematics, University of Rome ``Tor Vergata'', CUP E83C18000100006 and
the University of Rome ``Tor Vergata'' funding scheme ``Beyond Borders'', CUP E84I19002200005.
\end{acknowledgments}


\begin{thebibliography}{51}%
\makeatletter
\providecommand \@ifxundefined [1]{%
 \@ifx{#1\undefined}
}%
\providecommand \@ifnum [1]{%
 \ifnum #1\expandafter \@firstoftwo
 \else \expandafter \@secondoftwo
 \fi
}%
\providecommand \@ifx [1]{%
 \ifx #1\expandafter \@firstoftwo
 \else \expandafter \@secondoftwo
 \fi
}%
\providecommand \natexlab [1]{#1}%
\providecommand \enquote  [1]{``#1''}%
\providecommand \bibnamefont  [1]{#1}%
\providecommand \bibfnamefont [1]{#1}%
\providecommand \citenamefont [1]{#1}%
\providecommand \href@noop [0]{\@secondoftwo}%
\providecommand \href [0]{\begingroup \@sanitize@url \@href}%
\providecommand \@href[1]{\@@startlink{#1}\@@href}%
\providecommand \@@href[1]{\endgroup#1\@@endlink}%
\providecommand \@sanitize@url [0]{\catcode `\\12\catcode `\$12\catcode
  `\&12\catcode `\#12\catcode `\^12\catcode `\_12\catcode `\%12\relax}%
\providecommand \@@startlink[1]{}%
\providecommand \@@endlink[0]{}%
\providecommand \url  [0]{\begingroup\@sanitize@url \@url }%
\providecommand \@url [1]{\endgroup\@href {#1}{\urlprefix }}%
\providecommand \urlprefix  [0]{URL }%
\providecommand \Eprint [0]{\href }%
\providecommand \doibase [0]{https://doi.org/}%
\providecommand \selectlanguage [0]{\@gobble}%
\providecommand \bibinfo  [0]{\@secondoftwo}%
\providecommand \bibfield  [0]{\@secondoftwo}%
\providecommand \translation [1]{[#1]}%
\providecommand \BibitemOpen [0]{}%
\providecommand \bibitemStop [0]{}%
\providecommand \bibitemNoStop [0]{.\EOS\space}%
\providecommand \EOS [0]{\spacefactor3000\relax}%
\providecommand \BibitemShut  [1]{\csname bibitem#1\endcsname}%
\let\auto@bib@innerbib\@empty
\bibitem [{\citenamefont {{F}ern{\'a}ndez}\ \emph {et~al.}(1992)\citenamefont
  {{F}ern{\'a}ndez}, \citenamefont {{F}r{\"o}hlich},\ and\ \citenamefont
  {{S}okal}}]{FernandezRandomWalksCritical}%
  \BibitemOpen
  \bibfield  {author} {\bibinfo {author} {\bibfnamefont {R.}~\bibnamefont
  {{F}ern{\'a}ndez}}, \bibinfo {author} {\bibfnamefont {J.}~\bibnamefont
  {{F}r{\"o}hlich}},\ and\ \bibinfo {author} {\bibfnamefont {A.~D.}\
  \bibnamefont {{S}okal}},\ }\href {https://doi.org/10.1007/978-3-662-02866-7}
  {\emph {\bibinfo {title} {{Random walks, critical phenomena, and triviality
  in quantum field theory}}}},\ Texts and Monographs in Physics\ (\bibinfo
  {publisher} {{S}pringer, {B}erlin},\ \bibinfo {year} {1992})\BibitemShut
  {NoStop}%
\bibitem [{\citenamefont {Duerr}\ \emph {et~al.}(2008)\citenamefont {Duerr},
  \citenamefont {Fodor}, \citenamefont {Frison}, \citenamefont {Hoelbling},
  \citenamefont {Hoffmann}, \citenamefont {Katz}, \citenamefont {Krieg},
  \citenamefont {Kurth}, \citenamefont {Lellouch}, \citenamefont {Lippert},
  \citenamefont {Szabo},\ and\ \citenamefont
  {Vulvert}}]{DuerrAbInitioDetermination}%
  \BibitemOpen
  \bibfield  {author} {\bibinfo {author} {\bibfnamefont {S.}~\bibnamefont
  {Duerr}}, \bibinfo {author} {\bibfnamefont {Z.}~\bibnamefont {Fodor}},
  \bibinfo {author} {\bibfnamefont {J.}~\bibnamefont {Frison}}, \bibinfo
  {author} {\bibfnamefont {C.}~\bibnamefont {Hoelbling}}, \bibinfo {author}
  {\bibfnamefont {R.}~\bibnamefont {Hoffmann}}, \bibinfo {author}
  {\bibfnamefont {S.~D.}\ \bibnamefont {Katz}}, \bibinfo {author}
  {\bibfnamefont {S.}~\bibnamefont {Krieg}}, \bibinfo {author} {\bibfnamefont
  {T.}~\bibnamefont {Kurth}}, \bibinfo {author} {\bibfnamefont
  {L.}~\bibnamefont {Lellouch}}, \bibinfo {author} {\bibfnamefont
  {T.}~\bibnamefont {Lippert}}, \bibinfo {author} {\bibfnamefont {K.~K.}\
  \bibnamefont {Szabo}},\ and\ \bibinfo {author} {\bibfnamefont
  {G.}~\bibnamefont {Vulvert}},\ }\bibfield  {title} {\bibinfo {title}
  {{Ab-Initio Determination of Light Hadron Masses}},\ }\href
  {https://doi.org/10.1126/science.1163233} {\bibfield  {journal} {\bibinfo
  {journal} {Science}\ }\textbf {\bibinfo {volume} {322}},\ \bibinfo {pages}
  {1224} (\bibinfo {year} {2008})}\BibitemShut {NoStop}%
\bibitem [{\citenamefont {Jordan}\ \emph {et~al.}(2012)\citenamefont {Jordan},
  \citenamefont {Lee},\ and\ \citenamefont
  {Preskill}}]{JordanQuantumAlgorithmsFor}%
  \BibitemOpen
  \bibfield  {author} {\bibinfo {author} {\bibfnamefont {S.~P.}\ \bibnamefont
  {Jordan}}, \bibinfo {author} {\bibfnamefont {K.~S.~M.}\ \bibnamefont {Lee}},\
  and\ \bibinfo {author} {\bibfnamefont {J.}~\bibnamefont {Preskill}},\
  }\bibfield  {title} {\bibinfo {title} {{Quantum Algorithms for Quantum Field
  Theories}},\ }\href {https://doi.org/10.1126/science.1217069} {\bibfield
  {journal} {\bibinfo  {journal} {Science}\ }\textbf {\bibinfo {volume}
  {336}},\ \bibinfo {pages} {1130} (\bibinfo {year} {2012})},\ \Eprint
  {https://arxiv.org/abs/1111.3633} {1111.3633} \BibitemShut {NoStop}%
\bibitem [{\citenamefont {Glimm}\ and\ \citenamefont
  {Jaffe}(1985)}]{GlimmQuantumFieldTheory}%
  \BibitemOpen
  \bibfield  {author} {\bibinfo {author} {\bibfnamefont {J.}~\bibnamefont
  {Glimm}}\ and\ \bibinfo {author} {\bibfnamefont {A.}~\bibnamefont {Jaffe}},\
  }\href {https://doi.org/10.1007/978-1-4612-5158-3} {\emph {\bibinfo {title}
  {{Quantum Field Theory and Statistical Mechanics: Expositions}}}}\ (\bibinfo
  {publisher} {Birkhäuser, Boston},\ \bibinfo {year} {1985})\BibitemShut
  {NoStop}%
\bibitem [{\citenamefont {Jones}(2018{\natexlab{a}})}]{JonesANoGo}%
  \BibitemOpen
  \bibfield  {author} {\bibinfo {author} {\bibfnamefont {V.~F.~R.}\
  \bibnamefont {Jones}},\ }\bibfield  {title} {\bibinfo {title} {{A No-Go
  Theorem for the Continuum Limit of a Periodic Quantum Spin Chain}},\ }\href
  {https://doi.org/10.1007/s00220-017-2945-3} {\bibfield  {journal} {\bibinfo
  {journal} {Communications in Mathematical Physics}\ }\textbf {\bibinfo
  {volume} {357}},\ \bibinfo {pages} {295} (\bibinfo {year}
  {2018}{\natexlab{a}})}\BibitemShut {NoStop}%
\bibitem [{\citenamefont
  {Jones}(2018{\natexlab{b}})}]{JonesScaleInvariantTransfer}%
  \BibitemOpen
  \bibfield  {author} {\bibinfo {author} {\bibfnamefont {V.~F.~R.}\
  \bibnamefont {Jones}},\ }\bibfield  {title} {\bibinfo {title} {{Scale
  invariant transfer matrices and Hamiltionians}},\ }\href
  {https://doi.org/10.1088/1751-8121/aaa4dd} {\bibfield  {journal} {\bibinfo
  {journal} {Journal of Physics A: Mathematical and Theoretical}\ }\textbf
  {\bibinfo {volume} {51}},\ \bibinfo {pages} {104001} (\bibinfo {year}
  {2018}{\natexlab{b}})}\BibitemShut {NoStop}%
\bibitem [{\citenamefont {Kliesch}\ and\ \citenamefont
  {Koenig}(2020)}]{KlieschContinuumLimitsOf}%
  \BibitemOpen
  \bibfield  {author} {\bibinfo {author} {\bibfnamefont {A.}~\bibnamefont
  {Kliesch}}\ and\ \bibinfo {author} {\bibfnamefont {R.}~\bibnamefont
  {Koenig}},\ }\bibfield  {title} {\bibinfo {title} {{Continuum limits of
  homogeneous binary trees and the Thompson group}},\ }\href
  {https://doi.org/10.1103/PhysRevLett.124.010601} {\bibfield  {journal}
  {\bibinfo  {journal} {Physical Review Letters}\ }\textbf {\bibinfo {volume}
  {124}},\ \bibinfo {pages} {010601} (\bibinfo {year} {2020})}\BibitemShut
  {NoStop}%
\bibitem [{\citenamefont {Osborne}\ and\ \citenamefont
  {Stiegemann}(2019)}]{OsborneQuantumFieldsFor}%
  \BibitemOpen
  \bibfield  {author} {\bibinfo {author} {\bibfnamefont {T.~J.}\ \bibnamefont
  {Osborne}}\ and\ \bibinfo {author} {\bibfnamefont {D.~E.}\ \bibnamefont
  {Stiegemann}},\ }\bibfield  {title} {\bibinfo {title} {{Quantum fields for
  unitary representations of Thompson's groups F and T}},\ }\href
  {http://arxiv.org/abs/1903.00318v1} {\bibfield  {journal} {\bibinfo
  {journal} {Preprint, arXiv:1903.00318}\ } (\bibinfo {year}
  {2019})}\BibitemShut {NoStop}%
\bibitem [{\citenamefont {Kadanoff}(1966)}]{KadanoffScalingLawsFor}%
  \BibitemOpen
  \bibfield  {author} {\bibinfo {author} {\bibfnamefont {L.~P.}\ \bibnamefont
  {Kadanoff}},\ }\bibfield  {title} {\bibinfo {title} {{Scaling laws for Ising
  models near $T_{c}$}},\ }\href
  {https://doi.org/10.1103/PhysicsPhysiqueFizika.2.263} {\bibfield  {journal}
  {\bibinfo  {journal} {Physics Physique Fizika}\ }\textbf {\bibinfo {volume}
  {2}},\ \bibinfo {pages} {263} (\bibinfo {year} {1966})}\BibitemShut {NoStop}%
\bibitem [{\citenamefont {Wilson}(1975)}]{WilsonTheRenormalizationGroupKondo}%
  \BibitemOpen
  \bibfield  {author} {\bibinfo {author} {\bibfnamefont {K.~G.}\ \bibnamefont
  {Wilson}},\ }\bibfield  {title} {\bibinfo {title} {{The renormalization
  group: Critical phenomena and the Kondo problem}},\ }\href
  {https://doi.org/10.1103/RevModPhys.47.773} {\bibfield  {journal} {\bibinfo
  {journal} {Reviews of Modern Physics}\ }\textbf {\bibinfo {volume} {47}},\
  \bibinfo {pages} {773} (\bibinfo {year} {1975})}\BibitemShut {NoStop}%
\bibitem [{\citenamefont {Fisher}(1998)}]{FisherRenormalizationGroupTheory}%
  \BibitemOpen
  \bibfield  {author} {\bibinfo {author} {\bibfnamefont {M.~E.}\ \bibnamefont
  {Fisher}},\ }\bibfield  {title} {\bibinfo {title} {{Renormalization group
  theory: Its basis and formulation in statistical physics}},\ }\href
  {https://doi.org/10.1103/RevModPhys.70.653} {\bibfield  {journal} {\bibinfo
  {journal} {Reviews of Modern Physics}\ }\textbf {\bibinfo {volume} {70}},\
  \bibinfo {pages} {653} (\bibinfo {year} {1998})}\BibitemShut {NoStop}%
\bibitem [{\citenamefont {Brothier}\ and\ \citenamefont
  {Stottmeister}(2020)}]{BrothierConstructionsOfConformal}%
  \BibitemOpen
  \bibfield  {author} {\bibinfo {author} {\bibfnamefont {A.}~\bibnamefont
  {Brothier}}\ and\ \bibinfo {author} {\bibfnamefont {A.}~\bibnamefont
  {Stottmeister}},\ }\bibfield  {title} {\bibinfo {title} {{Operator-algebraic
  Construction of Gauge Theories and Jones’ Actions of Thompson’s
  Groups}},\ }\href {https://doi.org/10.1007/s00220-019-03603-4} {\bibfield
  {journal} {\bibinfo  {journal} {Communications in Mathematical Physics}\
  }\textbf {\bibinfo {volume} {376}},\ \bibinfo {pages} {841} (\bibinfo {year}
  {2020})},\ \Eprint {https://arxiv.org/abs/https://arxiv.org/abs/1901.04940}
  {https://arxiv.org/abs/1901.04940} \BibitemShut {NoStop}%
\bibitem [{\citenamefont {Brothier}\ and\ \citenamefont
  {Stottmeister}(2019)}]{BrothierAnOperatorAlgebraic}%
  \BibitemOpen
  \bibfield  {author} {\bibinfo {author} {\bibfnamefont {A.}~\bibnamefont
  {Brothier}}\ and\ \bibinfo {author} {\bibfnamefont {A.}~\bibnamefont
  {Stottmeister}},\ }\bibfield  {title} {\bibinfo {title} {{Canonical
  Quantization of 1+1-dimensional Yang-Mills Theory: An Operator Algebraic
  Approach}},\ }\href {http://arxiv.org/abs/1907.05549} {\bibfield  {journal}
  {\bibinfo  {journal} {arXiv: 1907.05549}\ } (\bibinfo {year}
  {2019})}\BibitemShut {NoStop}%
\bibitem [{\citenamefont {White}\ and\ \citenamefont
  {Noack}(1992)}]{WhiteRealSpaceQuantum1}%
  \BibitemOpen
  \bibfield  {author} {\bibinfo {author} {\bibfnamefont {S.~R.}\ \bibnamefont
  {White}}\ and\ \bibinfo {author} {\bibfnamefont {R.~M.}\ \bibnamefont
  {Noack}},\ }\bibfield  {title} {\bibinfo {title} {Real-space quantum
  renormalization groups},\ }\href
  {https://doi.org/10.1103/PhysRevLett.68.3487} {\bibfield  {journal} {\bibinfo
   {journal} {Physical Review Letters}\ }\textbf {\bibinfo {volume} {68}},\
  \bibinfo {pages} {3487} (\bibinfo {year} {1992})}\BibitemShut {NoStop}%
\bibitem [{\citenamefont {White}(1992)}]{WhiteDensityMatrixFormulation}%
  \BibitemOpen
  \bibfield  {author} {\bibinfo {author} {\bibfnamefont {S.~R.}\ \bibnamefont
  {White}},\ }\bibfield  {title} {\bibinfo {title} {Density matrix formulation
  for quantum renormalization groups},\ }\href
  {https://doi.org/10.1103/PhysRevLett.69.2863} {\bibfield  {journal} {\bibinfo
   {journal} {Physical review letters}\ }\textbf {\bibinfo {volume} {69}},\
  \bibinfo {pages} {2863} (\bibinfo {year} {1992})}\BibitemShut {NoStop}%
\bibitem [{\citenamefont {Morinelli}\ \emph {et~al.}(2021)\citenamefont
  {Morinelli}, \citenamefont {Morsella}, \citenamefont {Stottmeister},\ and\
  \citenamefont {Tanimoto}}]{MorinelliScalingLimitsOf}%
  \BibitemOpen
  \bibfield  {author} {\bibinfo {author} {\bibfnamefont {V.}~\bibnamefont
  {Morinelli}}, \bibinfo {author} {\bibfnamefont {G.}~\bibnamefont {Morsella}},
  \bibinfo {author} {\bibfnamefont {A.}~\bibnamefont {Stottmeister}},\ and\
  \bibinfo {author} {\bibfnamefont {Y.}~\bibnamefont {Tanimoto}},\ }\bibfield
  {title} {\bibinfo {title} {{Scaling limits of lattice quantum fields by
  wavelets}},\ }\bibfield  {journal} {\bibinfo  {journal} {Communications in
  Mathematical Physics}\ }\href {https://doi.org/10.1007/s00220-021-04152-5}
  {10.1007/s00220-021-04152-5} (\bibinfo {year} {2021}),\ \Eprint
  {https://arxiv.org/abs/http://arxiv.org/abs/2010.11121v1}
  {http://arxiv.org/abs/2010.11121v1} \BibitemShut {NoStop}%
\bibitem [{\citenamefont {Osborne}\ and\ \citenamefont
  {Stottmeister}(2021{\natexlab{a}})}]{OsborneCFTapprox}%
  \BibitemOpen
  \bibfield  {author} {\bibinfo {author} {\bibfnamefont {T.~J.}\ \bibnamefont
  {Osborne}}\ and\ \bibinfo {author} {\bibfnamefont {A.}~\bibnamefont
  {Stottmeister}},\ }\bibfield  {title} {\bibinfo {title} {{Conformal field
  theory from lattice fermions}},\ }\href {https://arxiv.org/abs/2107.13834}
  {\bibfield  {journal} {\bibinfo  {journal} {arXiv: 2107.13834}\ } (\bibinfo
  {year} {2021}{\natexlab{a}})}\BibitemShut {NoStop}%
\bibitem [{\citenamefont {Koo}\ and\ \citenamefont
  {Saleur}(1994)}]{KooRepresentationsOfThe}%
  \BibitemOpen
  \bibfield  {author} {\bibinfo {author} {\bibfnamefont {W.~M.}\ \bibnamefont
  {Koo}}\ and\ \bibinfo {author} {\bibfnamefont {H.}~\bibnamefont {Saleur}},\
  }\bibfield  {title} {\bibinfo {title} {{Representations of the Virasoro
  algebra from lattice models}},\ }\href
  {https://doi.org/10.1016/0550-3213(94)90018-3} {\bibfield  {journal}
  {\bibinfo  {journal} {Nuclear Physics B}\ }\textbf {\bibinfo {volume}
  {426}},\ \bibinfo {pages} {459} (\bibinfo {year} {1994})},\ \Eprint
  {https://arxiv.org/abs/arXiv preprint hep-th/9312156} {arXiv preprint
  hep-th/9312156} \BibitemShut {NoStop}%
\bibitem [{\citenamefont {Daubechies}(1992)}]{DaubechiesTenLecturesOn}%
  \BibitemOpen
  \bibfield  {author} {\bibinfo {author} {\bibfnamefont {I.}~\bibnamefont
  {Daubechies}},\ }\href {https://doi.org/10.1137/1.9781611970104} {\emph
  {\bibinfo {title} {{Ten Lectures on Wavelets}}}},\ \bibinfo {series}
  {CBMS-NSF Regional Conference Series in Applied Mathematics}, Vol.~\bibinfo
  {volume} {61}\ (\bibinfo  {publisher} {SIAM},\ \bibinfo {year}
  {1992})\BibitemShut {NoStop}%
\bibitem [{\citenamefont {Battle}(1999)}]{BattleWaveletsAndRenormalization}%
  \BibitemOpen
  \bibfield  {author} {\bibinfo {author} {\bibfnamefont {G.}~\bibnamefont
  {Battle}},\ }\href {https://doi.org/10.1142/3066} {\emph {\bibinfo {title}
  {{Wavelets and Renormalization}}}},\ \bibinfo {series} {Series in
  Approximations and Decompositions}, Vol.~\bibinfo {volume} {10}\ (\bibinfo
  {publisher} {World World Scientific Publishing Company},\ \bibinfo {year}
  {1999})\BibitemShut {NoStop}%
\bibitem [{\citenamefont {Brennen}\ \emph {et~al.}(2015)\citenamefont
  {Brennen}, \citenamefont {Rohde}, \citenamefont {Sanders},\ and\
  \citenamefont {Singh}}]{BrennenMultiscaleQuantumSimulation}%
  \BibitemOpen
  \bibfield  {author} {\bibinfo {author} {\bibfnamefont {G.~K.}\ \bibnamefont
  {Brennen}}, \bibinfo {author} {\bibfnamefont {P.}~\bibnamefont {Rohde}},
  \bibinfo {author} {\bibfnamefont {B.~C.}\ \bibnamefont {Sanders}},\ and\
  \bibinfo {author} {\bibfnamefont {S.}~\bibnamefont {Singh}},\ }\bibfield
  {title} {\bibinfo {title} {Multiscale quantum simulation of quantum field
  theory using wavelets},\ }\href {https://doi.org/10.1103/physreva.92.032315}
  {\bibfield  {journal} {\bibinfo  {journal} {Physical Review A: Atomic,
  Molecular, and Optical Physics}\ }\textbf {\bibinfo {volume} {92}},\ \bibinfo
  {pages} {032315} (\bibinfo {year} {2015})}\BibitemShut {NoStop}%
\bibitem [{\citenamefont {Evenbly}\ and\ \citenamefont
  {Vidal}(2010)}]{EvenblyEntanglementRenormalizationIn}%
  \BibitemOpen
  \bibfield  {author} {\bibinfo {author} {\bibfnamefont {G.}~\bibnamefont
  {Evenbly}}\ and\ \bibinfo {author} {\bibfnamefont {G.}~\bibnamefont
  {Vidal}},\ }\bibfield  {title} {\bibinfo {title} {{Entanglement
  renormalization in free bosonic systems: real-space versus momentum-space
  renormalization group transforms}},\ }\href
  {https://doi.org/10.1088/1367-2630/12/2/025007} {\bibfield  {journal}
  {\bibinfo  {journal} {New Journal of Physics}\ }\textbf {\bibinfo {volume}
  {12}},\ \bibinfo {pages} {025007} (\bibinfo {year} {2010})}\BibitemShut
  {NoStop}%
\bibitem [{\citenamefont {Zini}\ and\ \citenamefont
  {Wang}(2018)}]{ZiniConformalFieldTheories}%
  \BibitemOpen
  \bibfield  {author} {\bibinfo {author} {\bibfnamefont {M.~S.}\ \bibnamefont
  {Zini}}\ and\ \bibinfo {author} {\bibfnamefont {Z.}~\bibnamefont {Wang}},\
  }\bibfield  {title} {\bibinfo {title} {{Conformal Field Theories as Scaling
  Limit of Anyonic Chains}},\ }\href
  {https://doi.org/10.1007/s00220-018-3254-1} {\bibfield  {journal} {\bibinfo
  {journal} {Communications in Mathematical Physics}\ }\textbf {\bibinfo
  {volume} {363}},\ \bibinfo {pages} {877} (\bibinfo {year}
  {2018})}\BibitemShut {NoStop}%
\bibitem [{\citenamefont {White}(1993)}]{WhiteDensityMatrixAlgorithms}%
  \BibitemOpen
  \bibfield  {author} {\bibinfo {author} {\bibfnamefont {S.~R.}\ \bibnamefont
  {White}},\ }\bibfield  {title} {\bibinfo {title} {{Density-matrix algorithms
  for quantum renormalization groups}},\ }\href
  {https://doi.org/10.1103/PhysRevB.48.10345} {\bibfield  {journal} {\bibinfo
  {journal} {Physical Review B}\ }\textbf {\bibinfo {volume} {48}},\ \bibinfo
  {pages} {10345} (\bibinfo {year} {1993})}\BibitemShut {NoStop}%
\bibitem [{\citenamefont {Schollwöck}(2005)}]{SchollwoeckTheDensityMatrix}%
  \BibitemOpen
  \bibfield  {author} {\bibinfo {author} {\bibfnamefont {U.}~\bibnamefont
  {Schollwöck}},\ }\bibfield  {title} {\bibinfo {title} {The density-matrix
  renormalization group},\ }\href {https://doi.org/10.1103/revmodphys.77.259}
  {\bibfield  {journal} {\bibinfo  {journal} {Reviews of Modern Physics}\
  }\textbf {\bibinfo {volume} {77}},\ \bibinfo {pages} {259} (\bibinfo {year}
  {2005})}\BibitemShut {NoStop}%
\bibitem [{\citenamefont {{H}aag}(1996)}]{HaagLocalQuantumPhysics}%
  \BibitemOpen
  \bibfield  {author} {\bibinfo {author} {\bibfnamefont {R.}~\bibnamefont
  {{H}aag}},\ }\href {https://doi.org/10.1007/978-3-642-61458-3} {\emph
  {\bibinfo {title} {{L}ocal {Q}uantum {P}hysics: {F}ields, {P}articles,
  {A}lgebras}}},\ \bibinfo {edition} {2nd}\ ed.,\ Texts and Monographs in
  Physics\ (\bibinfo  {publisher} {{S}pringer {V}erlag},\ \bibinfo {year}
  {1996})\BibitemShut {NoStop}%
\bibitem [{\citenamefont {Lieb}\ and\ \citenamefont
  {Robinson}(1972)}]{LiebTheFiniteGroup}%
  \BibitemOpen
  \bibfield  {author} {\bibinfo {author} {\bibfnamefont {E.~H.}\ \bibnamefont
  {Lieb}}\ and\ \bibinfo {author} {\bibfnamefont {D.~W.}\ \bibnamefont
  {Robinson}},\ }\bibfield  {title} {\bibinfo {title} {The finite group
  velocity of quantum spin systems},\ }\href
  {https://doi.org/10.1007/BF01645779} {\bibfield  {journal} {\bibinfo
  {journal} {Communications in Mathematical Physics}\ }\textbf {\bibinfo
  {volume} {28}},\ \bibinfo {pages} {425} (\bibinfo {year} {1972})}\BibitemShut
  {NoStop}%
\bibitem [{\citenamefont {Cramer}\ \emph {et~al.}(2008)\citenamefont {Cramer},
  \citenamefont {Serafini},\ and\ \citenamefont
  {Eisert}}]{CramerLocalityOfDynamics}%
  \BibitemOpen
  \bibfield  {author} {\bibinfo {author} {\bibfnamefont {M.}~\bibnamefont
  {Cramer}}, \bibinfo {author} {\bibfnamefont {A.}~\bibnamefont {Serafini}},\
  and\ \bibinfo {author} {\bibfnamefont {J.}~\bibnamefont {Eisert}},\
  }\bibfield  {title} {\bibinfo {title} {Locality of dynamics in general
  harmonic quantum systems},\ }in\ \href
  {http://www.springer.com/birkhauser/mathematics/scuola normale
  superiore/book/978-88-7642-307-9} {\emph {\bibinfo {booktitle} {Quantum
  information and many body quantum systems}}},\ \bibinfo {series}
  {Publications of the Scuola Normale Superiore, CRM}, Vol.~\bibinfo {volume}
  {8},\ \bibinfo {editor} {edited by\ \bibinfo {editor} {\bibfnamefont
  {M.}~\bibnamefont {Ericsson}}\ and\ \bibinfo {editor} {\bibfnamefont
  {S.}~\bibnamefont {Montangero}}}\ (\bibinfo {year} {2008})\ pp.\ \bibinfo
  {pages} {51--72},\ \Eprint
  {https://arxiv.org/abs/http://arxiv.org/abs/0803.0890}
  {http://arxiv.org/abs/0803.0890} \BibitemShut {NoStop}%
\bibitem [{\citenamefont {Osborne}(2019)}]{OsborneContinuumLimitsOf}%
  \BibitemOpen
  \bibfield  {author} {\bibinfo {author} {\bibfnamefont {T.~J.}\ \bibnamefont
  {Osborne}},\ }\bibfield  {title} {\bibinfo {title} {{Continuum Limits of
  Quantum Lattice Systems}},\ }\href {http://arxiv.org/abs/1901.06124v1}
  {\bibfield  {journal} {\bibinfo  {journal} {Preprint, arXiv:1901.06124}\ }
  (\bibinfo {year} {2019})},\ \Eprint
  {https://arxiv.org/abs/http://arxiv.org/abs/1901.06124v1}
  {http://arxiv.org/abs/1901.06124v1} \BibitemShut {NoStop}%
\bibitem [{\citenamefont {Nachtergaele}\ \emph {et~al.}(2009)\citenamefont
  {Nachtergaele}, \citenamefont {Raz}, \citenamefont {Schlein},\ and\
  \citenamefont {Sims}}]{NachtergaeleLRBoundsHarmonic}%
  \BibitemOpen
  \bibfield  {author} {\bibinfo {author} {\bibfnamefont {B.}~\bibnamefont
  {Nachtergaele}}, \bibinfo {author} {\bibfnamefont {H.}~\bibnamefont {Raz}},
  \bibinfo {author} {\bibfnamefont {B.}~\bibnamefont {Schlein}},\ and\ \bibinfo
  {author} {\bibfnamefont {R.}~\bibnamefont {Sims}},\ }\bibfield  {title}
  {\bibinfo {title} {{Lieb-Robinson bounds for harmonic and anharmonic lattice
  systems}},\ }\href {https://doi.org/10.1007/s00220-008-0630-2} {\bibfield
  {journal} {\bibinfo  {journal} {Communications in Mathematical Physics}\
  }\textbf {\bibinfo {volume} {286}},\ \bibinfo {pages} {1073} (\bibinfo {year}
  {2009})}\BibitemShut {NoStop}%
\bibitem [{\citenamefont {Nachtergaele}\ \emph {et~al.}(2019)\citenamefont
  {Nachtergaele}, \citenamefont {Sims},\ and\ \citenamefont
  {Young}}]{NachtergaeleQuasiLocalityBounds}%
  \BibitemOpen
  \bibfield  {author} {\bibinfo {author} {\bibfnamefont {B.}~\bibnamefont
  {Nachtergaele}}, \bibinfo {author} {\bibfnamefont {R.}~\bibnamefont {Sims}},\
  and\ \bibinfo {author} {\bibfnamefont {A.}~\bibnamefont {Young}},\ }\bibfield
   {title} {\bibinfo {title} {{Quasi-locality bounds for quantum lattice
  systems. I. Lieb-Robinson bounds, quasi-local maps, and spectral flow
  automorphisms}},\ }\href {https://doi.org/10.1063/1.5095769} {\bibfield
  {journal} {\bibinfo  {journal} {Journal of Mathematical Physics}\ }\textbf
  {\bibinfo {volume} {60}},\ \bibinfo {pages} {061101} (\bibinfo {year}
  {2019})}\BibitemShut {NoStop}%
\bibitem [{\citenamefont {Cirac}\ and\ \citenamefont
  {Verstraete}(2009)}]{CiracRenormalizationAndTensor}%
  \BibitemOpen
  \bibfield  {author} {\bibinfo {author} {\bibfnamefont {J.~I.}\ \bibnamefont
  {Cirac}}\ and\ \bibinfo {author} {\bibfnamefont {F.}~\bibnamefont
  {Verstraete}},\ }\bibfield  {title} {\bibinfo {title} {Renormalization and
  tensor product states in spin chains and lattices},\ }\href
  {https://doi.org/10.1088/1751-8113/42/50/504004} {\bibfield  {journal}
  {\bibinfo  {journal} {Journal of Physics A: Mathematical and Theoretical}\
  }\textbf {\bibinfo {volume} {42}},\ \bibinfo {pages} {504004} (\bibinfo
  {year} {2009})}\BibitemShut {NoStop}%
\bibitem [{\citenamefont {Vidal}(2008)}]{VidalAClassOf}%
  \BibitemOpen
  \bibfield  {author} {\bibinfo {author} {\bibfnamefont {G.}~\bibnamefont
  {Vidal}},\ }\bibfield  {title} {\bibinfo {title} {{A Class of Quantum
  Many-Body States That Can Be Efficiently Simulated}},\ }\href
  {https://doi.org/10.1103/physrevlett.101.110501} {\bibfield  {journal}
  {\bibinfo  {journal} {Physical Review Letters}\ }\textbf {\bibinfo {volume}
  {101}},\ \bibinfo {pages} {110501} (\bibinfo {year} {2008})}\BibitemShut
  {NoStop}%
\bibitem [{\citenamefont {Evenbly}\ and\ \citenamefont
  {Vidal}(2009)}]{EvenblyAlgorithmsForEntanglement}%
  \BibitemOpen
  \bibfield  {author} {\bibinfo {author} {\bibfnamefont {G.}~\bibnamefont
  {Evenbly}}\ and\ \bibinfo {author} {\bibfnamefont {G.}~\bibnamefont
  {Vidal}},\ }\bibfield  {title} {\bibinfo {title} {Algorithms for entanglement
  renormalization},\ }\href {https://doi.org/10.1103/physrevb.79.144108}
  {\bibfield  {journal} {\bibinfo  {journal} {Physical Review B: Condensed
  Matter and Materials Physics}\ }\textbf {\bibinfo {volume} {79}},\ \bibinfo
  {pages} {144108} (\bibinfo {year} {2009})}\BibitemShut {NoStop}%
\bibitem [{\citenamefont {Pfeifer}\ \emph {et~al.}(2009)\citenamefont
  {Pfeifer}, \citenamefont {Evenbly},\ and\ \citenamefont
  {Vidal}}]{PfeiferEntanglementRenormalizationScale}%
  \BibitemOpen
  \bibfield  {author} {\bibinfo {author} {\bibfnamefont {R.~N.~C.}\
  \bibnamefont {Pfeifer}}, \bibinfo {author} {\bibfnamefont {G.}~\bibnamefont
  {Evenbly}},\ and\ \bibinfo {author} {\bibfnamefont {G.}~\bibnamefont
  {Vidal}},\ }\bibfield  {title} {\bibinfo {title} {Entanglement
  renormalization, scale invariance, and quantum criticality},\ }\href
  {https://doi.org/10.1103/physreva.79.040301} {\bibfield  {journal} {\bibinfo
  {journal} {Physical Review A: Atomic, Molecular, and Optical Physics}\
  }\textbf {\bibinfo {volume} {79}},\ \bibinfo {pages} {040301} (\bibinfo
  {year} {2009})}\BibitemShut {NoStop}%
\bibitem [{\citenamefont {Evenbly}\ and\ \citenamefont
  {White}(2016)}]{EvenblyEntanglementRenormalizationAnd}%
  \BibitemOpen
  \bibfield  {author} {\bibinfo {author} {\bibfnamefont {G.}~\bibnamefont
  {Evenbly}}\ and\ \bibinfo {author} {\bibfnamefont {S.~R.}\ \bibnamefont
  {White}},\ }\bibfield  {title} {\bibinfo {title} {{Entanglement
  Renormalization and Wavelets}},\ }\href
  {https://doi.org/10.1103/physrevlett.116.140403} {\bibfield  {journal}
  {\bibinfo  {journal} {Physical Review Letters}\ }\textbf {\bibinfo {volume}
  {116}},\ \bibinfo {pages} {140403} (\bibinfo {year} {2016})}\BibitemShut
  {NoStop}%
\bibitem [{\citenamefont {Evenbly}\ and\ \citenamefont
  {White}(2018)}]{EvenblyRepresentationAndDesign}%
  \BibitemOpen
  \bibfield  {author} {\bibinfo {author} {\bibfnamefont {G.}~\bibnamefont
  {Evenbly}}\ and\ \bibinfo {author} {\bibfnamefont {S.~R.}\ \bibnamefont
  {White}},\ }\bibfield  {title} {\bibinfo {title} {Representation and design
  of wavelets using unitary circuits},\ }\href
  {https://doi.org/10.1103/physreva.97.052314} {\bibfield  {journal} {\bibinfo
  {journal} {Physical Review A: Atomic, Molecular, and Optical Physics}\
  }\textbf {\bibinfo {volume} {97}},\ \bibinfo {pages} {052314} (\bibinfo
  {year} {2018})}\BibitemShut {NoStop}%
\bibitem [{\citenamefont {Milsted}\ and\ \citenamefont
  {Osborne}(2018)}]{MilstedQuantumYangMills}%
  \BibitemOpen
  \bibfield  {author} {\bibinfo {author} {\bibfnamefont {A.}~\bibnamefont
  {Milsted}}\ and\ \bibinfo {author} {\bibfnamefont {T.~J.}\ \bibnamefont
  {Osborne}},\ }\bibfield  {title} {\bibinfo {title} {{Quantum Yang-Mills
  theory: An overview of a program}},\ }\href
  {https://doi.org/10.1103/physrevd.98.014505} {\bibfield  {journal} {\bibinfo
  {journal} {Physical Review D: Particles and Fields}\ }\textbf {\bibinfo
  {volume} {98}},\ \bibinfo {pages} {014505} (\bibinfo {year}
  {2018})}\BibitemShut {NoStop}%
\bibitem [{\citenamefont {{K}ijowski}(1977)}]{KijowskiSymplecticGeometryAnd}%
  \BibitemOpen
  \bibfield  {author} {\bibinfo {author} {\bibfnamefont {J.}~\bibnamefont
  {{K}ijowski}},\ }\bibfield  {title} {\bibinfo {title} {{S}ymplectic geometry
  and second quantization},\ }\href@noop {} {\bibfield  {journal} {\bibinfo
  {journal} {{R}eports on {M}athematical {P}hysics}\ }\textbf {\bibinfo
  {volume} {11}},\ \bibinfo {pages} {97} (\bibinfo {year} {1977})}\BibitemShut
  {NoStop}%
\bibitem [{\citenamefont {Kijowski}\ and\ \citenamefont
  {Oko{\l}{\'o}w}(2017)}]{KijowskiAModificationOf}%
  \BibitemOpen
  \bibfield  {author} {\bibinfo {author} {\bibfnamefont {J.}~\bibnamefont
  {Kijowski}}\ and\ \bibinfo {author} {\bibfnamefont {A.}~\bibnamefont
  {Oko{\l}{\'o}w}},\ }\bibfield  {title} {\bibinfo {title} {A modification of
  the projective construction of quantum states for field theories},\ }\href
  {https://doi.org/10.1063/1.4989550} {\bibfield  {journal} {\bibinfo
  {journal} {Journal of Mathematical Physics}\ }\textbf {\bibinfo {volume}
  {58}},\ \bibinfo {pages} {062303} (\bibinfo {year} {2017})}\BibitemShut
  {NoStop}%
\bibitem [{\citenamefont {Meyer}(1989)}]{MeyerWaveletsAndOperators}%
  \BibitemOpen
  \bibfield  {author} {\bibinfo {author} {\bibfnamefont {Y.}~\bibnamefont
  {Meyer}},\ }\href {https://doi.org/10.1017/CBO9780511623820} {\emph {\bibinfo
  {title} {Wavelets and operators}}},\ \bibinfo {series} {Cambridge studies in
  advanced mathematics}, Vol.~\bibinfo {volume} {37}\ (\bibinfo  {publisher}
  {Cambridge University Press},\ \bibinfo {year} {1989})\BibitemShut {NoStop}%
\bibitem [{\citenamefont
  {Mallat}(1989)}]{MallatMultiresolutionApproximationsAnd}%
  \BibitemOpen
  \bibfield  {author} {\bibinfo {author} {\bibfnamefont {S.~G.}\ \bibnamefont
  {Mallat}},\ }\bibfield  {title} {\bibinfo {title} {{Multiresolution
  approximations and wavelet orthonormal bases of $L^{2}(\mathds{R})$}},\
  }\href {https://doi.org/10.1090/s0002-9947-1989-1008470-5} {\bibfield
  {journal} {\bibinfo  {journal} {{Transactions of the American Mathematical
  Society}}\ }\textbf {\bibinfo {volume} {315}},\ \bibinfo {pages} {69}
  (\bibinfo {year} {1989})}\BibitemShut {NoStop}%
\bibitem [{\citenamefont {Meyer}(1987)}]{MeyerPrincipeDIncertitude}%
  \BibitemOpen
  \bibfield  {author} {\bibinfo {author} {\bibfnamefont {Y.}~\bibnamefont
  {Meyer}},\ }\bibfield  {title} {\bibinfo {title} {Principe d'incertitude,
  bases hilbertiennes et algebres d'operateurs},\ }in\ \href
  {http://www.numdam.org/item/SB_1985-1986__28__209_0} {\emph {\bibinfo
  {booktitle} {{S\'eminaire Bourbaki : Volume 1985/86, Expos\'es 651-668}}}},\
  \bibinfo {series and number} {\bibinfo {series} {Ast\'erisque}\ No.\ \bibinfo
  {number} {145--146}}\ (\bibinfo  {publisher} {{Soci\'et\'e math\'ematique de
  France}},\ \bibinfo {year} {1987})\ pp.\ \bibinfo {pages}
  {209--223}\BibitemShut {NoStop}%
\bibitem [{\citenamefont {Haegeman}\ \emph {et~al.}(2018)\citenamefont
  {Haegeman}, \citenamefont {Swingle}, \citenamefont {Walter}, \citenamefont
  {Cotler}, \citenamefont {Evenbly},\ and\ \citenamefont
  {Scholz}}]{HaegemanRigorousFreeFermion}%
  \BibitemOpen
  \bibfield  {author} {\bibinfo {author} {\bibfnamefont {J.}~\bibnamefont
  {Haegeman}}, \bibinfo {author} {\bibfnamefont {B.}~\bibnamefont {Swingle}},
  \bibinfo {author} {\bibfnamefont {M.}~\bibnamefont {Walter}}, \bibinfo
  {author} {\bibfnamefont {J.}~\bibnamefont {Cotler}}, \bibinfo {author}
  {\bibfnamefont {G.}~\bibnamefont {Evenbly}},\ and\ \bibinfo {author}
  {\bibfnamefont {V.~B.}\ \bibnamefont {Scholz}},\ }\bibfield  {title}
  {\bibinfo {title} {{Rigorous free-fermion entanglement renormalization from
  wavelet theory}},\ }\href {https://doi.org/10.1103/PhysRevX.8.011003}
  {\bibfield  {journal} {\bibinfo  {journal} {Physical Review X}\ }\textbf
  {\bibinfo {volume} {8}},\ \bibinfo {pages} {011003} (\bibinfo {year}
  {2018})}\BibitemShut {NoStop}%
\bibitem [{\citenamefont {Witteveen}\ \emph {et~al.}(2019)\citenamefont
  {Witteveen}, \citenamefont {Scholz}, \citenamefont {Swingle},\ and\
  \citenamefont {Walter}}]{WitteveenQuantumCircuitApproximations}%
  \BibitemOpen
  \bibfield  {author} {\bibinfo {author} {\bibfnamefont {F.}~\bibnamefont
  {Witteveen}}, \bibinfo {author} {\bibfnamefont {V.}~\bibnamefont {Scholz}},
  \bibinfo {author} {\bibfnamefont {B.}~\bibnamefont {Swingle}},\ and\ \bibinfo
  {author} {\bibfnamefont {M.}~\bibnamefont {Walter}},\ }\bibfield  {title}
  {\bibinfo {title} {{Quantum circuit approximations and entanglement
  renormalization for the Dirac field in 1+1 dimensions}},\ }\href
  {https://arxiv.org/pdf/1905.08821v1} {\bibfield  {journal} {\bibinfo
  {journal} {Preprint, arXiv: 1905.08821}\ } (\bibinfo {year} {2019})},\
  \Eprint {https://arxiv.org/abs/1905.08821v1} {1905.08821v1} \BibitemShut
  {NoStop}%
\bibitem [{\citenamefont {Witteveen}\ and\ \citenamefont
  {Walter}(2021)}]{WitteveenWaveletConstructionOf}%
  \BibitemOpen
  \bibfield  {author} {\bibinfo {author} {\bibfnamefont {F.}~\bibnamefont
  {Witteveen}}\ and\ \bibinfo {author} {\bibfnamefont {M.}~\bibnamefont
  {Walter}},\ }\bibfield  {title} {\bibinfo {title} {Wavelet construction of
  bosonic entanglement renormalization circuits},\ }\href
  {https://doi.org/10.21468/SciPostPhys.10.6.143} {\bibfield  {journal}
  {\bibinfo  {journal} {SciPost Phys.}\ }\textbf {\bibinfo {volume} {10}},\
  \bibinfo {pages} {143} (\bibinfo {year} {2021})},\ \Eprint
  {https://arxiv.org/abs/https://arxiv.org/pdf/2004.11952v1}
  {https://arxiv.org/pdf/2004.11952v1} \BibitemShut {NoStop}%
\bibitem [{\citenamefont {Osborne}\ and\ \citenamefont
  {Stottmeister}(2021{\natexlab{b}})}]{OsborneCFTsim}%
  \BibitemOpen
  \bibfield  {author} {\bibinfo {author} {\bibfnamefont {T.~J.}\ \bibnamefont
  {Osborne}}\ and\ \bibinfo {author} {\bibfnamefont {A.}~\bibnamefont
  {Stottmeister}},\ }\bibfield  {title} {\bibinfo {title} {{Quantum Simulation
  of Conformal Field Theory}},\ }\href {https://arxiv.org/abs/2109.14214}
  {\bibfield  {journal} {\bibinfo  {journal} {arXiv: 2109.14214}\ } (\bibinfo
  {year} {2021}{\natexlab{b}})}\BibitemShut {NoStop}%
\bibitem [{\citenamefont {Borgs}(1988)}]{BorgsConfinementDeconfinementAnd}%
  \BibitemOpen
  \bibfield  {author} {\bibinfo {author} {\bibfnamefont {C.}~\bibnamefont
  {Borgs}},\ }\bibfield  {title} {\bibinfo {title} {{Confinement, deconfinement
  and freezing in lattice Yang-Mills theories with continuous time}},\ }\href
  {https://doi.org/10.1007/bf01225259} {\bibfield  {journal} {\bibinfo
  {journal} {Communications in Mathematical Physics}\ }\textbf {\bibinfo
  {volume} {116}},\ \bibinfo {pages} {309} (\bibinfo {year}
  {1988})}\BibitemShut {NoStop}%
\bibitem [{\citenamefont {Brydges}(2009)}]{BrydgesLecturesOnThe}%
  \BibitemOpen
  \bibfield  {author} {\bibinfo {author} {\bibfnamefont {D.}~\bibnamefont
  {Brydges}},\ }\bibinfo {title} {Lectures on the renormalisation group}\
  (\bibinfo  {publisher} {American Mathematical Society},\ \bibinfo {year}
  {2009})\ Chap.~\bibinfo {chapter} {1}, pp.\ \bibinfo {pages}
  {7--94}\BibitemShut {NoStop}%
\bibitem [{\citenamefont {Glimm}\ and\ \citenamefont
  {Jaffe}(1968)}]{GlimmJaffe1}%
  \BibitemOpen
  \bibfield  {author} {\bibinfo {author} {\bibfnamefont {J.}~\bibnamefont
  {Glimm}}\ and\ \bibinfo {author} {\bibfnamefont {A.}~\bibnamefont {Jaffe}},\
  }\bibfield  {title} {\bibinfo {title} {{A $\lambda\phi^{4}_{2}$ Quantum Field
  Theory without Cutoffs. I}},\ }\href
  {https://doi.org/10.1103/PhysRev.176.1945} {\bibfield  {journal} {\bibinfo
  {journal} {Physical Review}\ }\textbf {\bibinfo {volume} {176}},\ \bibinfo
  {pages} {1945} (\bibinfo {year} {1968})}\BibitemShut {NoStop}%
\bibitem [{\citenamefont {Glimm}\ and\ \citenamefont
  {Jaffe}(1970)}]{GlimmJaffe3}%
  \BibitemOpen
  \bibfield  {author} {\bibinfo {author} {\bibfnamefont {J.}~\bibnamefont
  {Glimm}}\ and\ \bibinfo {author} {\bibfnamefont {A.}~\bibnamefont {Jaffe}},\
  }\bibfield  {title} {\bibinfo {title} {{The $\lambda\phi^{4}_{2}$ quantum
  field theory without cutoffs, III. The physical vacuum}},\ }\href
  {https://doi.org/10.1007/BF02392335} {\bibfield  {journal} {\bibinfo
  {journal} {{Acta Mathematica}}\ }\textbf {\bibinfo {volume} {125}},\ \bibinfo
  {pages} {203} (\bibinfo {year} {1970})}\BibitemShut {NoStop}%
\end{thebibliography}

%

\end{document}